\documentclass{aa}  
\usepackage{graphicx}
\usepackage{txfonts}

\usepackage[breaklinks=true]{hyperref}
\usepackage{natbib}
\bibpunct{(}{)}{;}{a}{}{,} 

\usepackage[export]{adjustbox}
\usepackage{textcomp}

\def\aap{A\&A}
\def\apjl{ApJL}
\def\apjs{ApJS}

\def\orcid#1{\kern .08em\href{https://orcid.org/#1}{\includegraphics[keepaspectratio,width=0.7em]{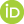}}}

\DeclareRobustCommand{\VAN}[3]{#2}
\let\VANthebibliography\thebibliography
\def\thebibliography{\DeclareRobustCommand{\VAN}[3]{##3}\VANthebibliography}
\begin{document}

   \title{Effects of the internal temperature on vertical mixing and on cloud structures in Ultra Hot Jupiters}

   \subtitle{}

   \author{Pascal A. Noti\orcid{0000-0002-8012-3400}
          \inst{1}\fnmsep\inst{2}\fnmsep\thanks{\email{pascal-andreas.noti@unibe.ch}}
          \and
          Elspeth K. H. Lee\orcid{0000-0002-3052-7116}
          \inst{1}
          }

   \institute{Center for Space and Habitability, Universit\"at Bern,
              Gesellschaftsstrasse 6, CH-3012 Bern, Switzerland\\
         \and
             Physikalisches Institut, Universität Bern,
              Sidlerstrasse 5, CH-3012 Bern, Switzerland\\
             }

   \date{Received 7 August 2024; Accepted 23 September 2024}

   \authorrunning{Noti et al. 2024}

 
  \abstract
   {The vertical mixing in hot Jupiter atmospheres plays a critical role in the formation and spacial distribution of cloud particles in their atmospheres.
   This affects the observed spectra of a planet through cloud opacity, which can be influenced by the degree of cold trapping of refractory species in the deep atmosphere.}
   {We aim to isolate the effects of the internal temperature on the mixing efficiency in the atmospheres of Ultra Hot Jupiters (UHJ) and the spacial distribution of cloud particles across the globe.}
   {We couple a simplified tracer based cloud model, picket fence radiative-transfer scheme and mixing length theory to the Exo-FMS general circulation model. 
   We run the model for five different internal temperatures at typical UHJ atmosphere system parameters.}
   {Our results show the convective eddy diffusion coefficient remains low throughout the vast majority of the atmosphere, with mixing dominated by advective flows. However, some regions can show convective mixing in the upper atmosphere for colder interior temperatures. The vertical extent of the clouds is reduced as the internal temperature is increased. Additionally, a global cloud layer gets formed below the radiative-convective boundary (RCB) in the cooler cases.}
   {Convection is generally strongly inhibited in UHJ atmospheres above the RCB due to their strong irradiation. Convective mixing plays a minor role compared to advective mixing in keeping cloud particles aloft in ultra hot Jupiters with warm interiors. Higher vertical turbulent heat fluxes and the advection of potential temperature inhibit convection in warmer interiors. Our results suggest isolated upper atmosphere regions above cold interiors may exhibit strong convective mixing in isolated regions around Rossby gyres, allowing aerosols to be better retained in these areas.}

   \keywords{Planets and satellites: atmospheres, Planets and satellites: gaseous planets, Planets and satellites: interiors, Hydrodynamics, Turbulence, Convection}

   \maketitle
%

\section{Introduction}


Our solar system planets differ from the intensely irradiated hot Jupiter exoplanets \citep{Guillot1996ApJ,Sager1998ApJ,Marley1999ApJ}. The high irradiation in combination with several factors such as chemical compositions, thermochemical pathways and various cloud species make exoplanets very complex. For instance, clouds lead to changes in the thermal and dynamical state of solar system giants as e.g. the Galileo probe measured while falling into a relatively hot and dry region of the Jovian atmosphere \citep{Showman2000Sci...289.1737S}. Such impact on the thermal and dynamical state by clouds is assumed as well in hot Jupiters (HJ hereafter) \citep{Roman2021,deitrick2022, Komacek2022PatchyClouds}. The high irradiation on the dayside lead to a radiative atmosphere to much greater depth \citep{Guillot&Showman2002A&A, showman&guillot2002, Sudarsky2003ApJ, thorngren2019}.

The radiative–convective boundary (RCB hereafter) plays a crucial role in modelling HJs and Ultra Hot Jupiters (UHJ hereafter) \citep{thorngren2019}, which depend on the internal irradiation temperature, $T_{int}$. Higher $T_{int}$ leads to a shallower RCB. A shallower RCB lead to significant changes in the atmospheric structure, dynamics, interpretation of atmospheric spectra, and the effect of deep cold traps on cloud formation  \citep{thorngren2019}. Lower $T_{int}$ in HJ can cold trap many chemical species and keep them in the deeper atmosphere \citep{parmentier2016}. The efficiency of the cold trap depends highly on the deep thermal structures and the strength of the vertical mixing in the deep atmosphere \citep{Spiegel2009ApJ, Powell2018}. Therefore, the deep temperature structure play a crucial role for convection as well as the related vertical mixing and cloud formation.

Since \citet{thorngren2019} and \cite{Sarkis2021A&A...645A..79S} showed that deep atmosphere of HJs and UHJs has likely higher temperatures than assumed in previous studies, the strength of the vertical mixing might change with higher temperatures in deep atmospheres. Although estimations of the vertical mixing has been formulated \citep{Lewis2010, Moses2011, Parmentier2013}, these estimations and the vertical extent of the clouds have already been questioned in recent studies like in \citet{Roman2021} and \citet{Malsky2024ApJ}.

Vertical transport and mixing counter the gravitational settling of condensates and have a crucial influence on the extent of clouds   \citep{Powell2018, Helling2019AREPS}. Cloud particles keep suspended in significant abundances only if the diffusion coefficient, $K_{zz}$ \lbrack m\textsuperscript{2}s \textsuperscript{-1}\rbrack \:, surpasses a critical value \citep{Parmentier2013}. The critical value depends highly on the particle size of the condensates.
The advective (effective) $K_{zz}$ can be retrieved from averaged vertical flux produced by the dynamics \citep[see][p.90]{Chamberlain1987tpaa.book.....C} as
\begin{equation}
K_{zz}=\frac{\langle \rho \chi w\rangle_h}{\langle \rho \frac{\partial \chi  }{\partial z}\rangle_h},
	\label{eq:Kzz_chamberlain}
\end{equation}
where $\rho$ \lbrack kgm\textsuperscript{-3}\rbrack \:is the density, $\chi$  denotes the mole fraction of the chemical species, $w$ \lbrack ms\textsuperscript{-1}\rbrack \:represents the vertical wind speed, and $\langle \rangle_h$ horizontal average along isobars across the planet.

Several previous studies have estimated the vertical diffusion coefficient for HJ \citep{cooper2006,showman2008,showman2009, Showman2013ApJ, Lewis2010, heng2011b, Moses2011, Parmentier2013}. 
In a first approach, \citet{cooper2006} made an estimation of $K_{zz}$ as the product of a root-mean-square vertical velocity. They extracted vertical velocity from their 3D Global Circulation Models (GCM hereafter) and put this product in relation to vertical length scale following the formulation of \citet{Smith1998Icar}. \citet{Lewis2010} and \citet{Moses2011} took a similar path the implementation of an atmospheric scale height instead of the vertical length scale. They defined the vertical advective $K_{zz}$ as
\begin{equation}
K_{zz}=H(z) \sqrt{\langle w^2\rangle_h},
	\label{eq:Kzz_Lewis_Moses}
\end{equation}
where $H(z)$ \lbrack m\rbrack \:stands for the scale height calculated with the average of the temperatures at the same pressure. These approach provide rough estimations, although \citet{cooper2006} showed how well their estimation of advective $K_{zz}$ works in 1D models. Later, \citet{heng2011b} developed an approach based on the magnitude of the Eulerian mean stream-function. They used this stream-function to derive the vertical mixing coefficient from the strength of the vertical motions. Their approach leads to $K_{zz}\sim 10^6$  ms\textsuperscript{-2}. Indeed, Eulerian-mean velocities can poorly describe tracer advection, since large-scale eddies mostly dominate over Eulerian-mean circulation regarding mixing \citep{Andrews1987book, Parmentier2013}. Estimating the mixing of tracers across isobars relies on the correlations between eddy tracer abundance and eddy vertical velocity \citep{Zhang2018ApJ...866....1Z, Zhang2018ApJ...866....2Z}. Later, \citet{Parmentier2013} tried a another approach with the flux gradient relation. Their 1D fit estimates $K_{zz}$ based on different tracer fields from GCM simulation of HD-209458b. Their specific fit estimates the vertical advective $K_{zz}$ as
\begin{equation}
K_{zz}=5\cdot10^4\sqrt{\frac{p}{10^5}},
	\label{eq:Kzz_Parmentier}
\end{equation}
where $p$ \lbrack Pa\rbrack \:is the pressure of the gas.
Another approach to estimate the vertical mixing based on the mixing length theory (MLT) \citep[e.g.][]{Joyce2023}. The MLT allows to capture the net large scale effects of convection onto the atmosphere by the convective diffusion coefficient \citep{Marley2015}. Only a few cloud related studies \citep[e.g.][]{Lee2024MNRAS.529.2686L} have used this recent approach to quantify vertical mixing.

UHJs are gas giant planets which orbit very closely their host stars within a few days. Therefore, they receive extremly high stellar irradiation which lead to full-redistribution equilibrium
temperatures beyond 2200 K. The physics on their ultra hot dayside and cooler nightside are in transition between late-type stars and the coolder HJ. \citep{Tan2024MNRAS}

The transition between stellar and HJ atmosphere makes UHJs interesting candidates to study convection, cloud structures and vertical mixing. Moreover, the strong irradiation on UHJs inhibits convection in upper atmosphere \citep{Parmentier2013}. Additionally, UHJ are expected to have higher $T_{int}$ than other planets, especially on the dayside. This high temperature suggest low heat flow likely caused by magnetic drag \citep{Maxted2013MNRAS.428.2645M}. Higher $T_{int}$ can lead to higher vertical mixing and dispersion of heavier elements.

The UHJ are the best targets for thermal emission measurements among the hotter planets because their contrast to the host star makes them easier to observe than cooler HJ. The UHJ observed so far have weaker spectral features in the 1-2 {\textmu}m range than cooler planets. For instance for WASP-121b, the weaker spectral features are generated by a combination of thermal dissociation which creates a vertical gradient in molecular abundances, and because of the H- absorption at wavelengths shorter than 1.4 {\textmu}m. This thermal dissociation changes the abundances of all spectrally important molecules, except CO. The change in the abundances create a large vertical gradients which differ among molecules. This changes the ratio between molecules vertically. For instance, the abundance ratio between Na and H2O increases with decreasing pressure. Consequently, Na absorbs the stellar light that much that thermal inversions appear on the dayside, even in absence of TiO and VO, and at solar composition. \citep{Parmentier2018A&A...617A.110P}

Recent observations of WASP-18b proved the thermal dissociation. \citet{Coulombe2023Natur.620..292C} revealed three water emission features and evidence for optical opacity, possibly caused by H-, TiO and VO. Their models fits to these observations need thermal inversion, thermal dissociation of molecules, a solar heavy-element abundance ('metallicity', $M/H=1.03$ times solar) and a carbon-to-oxygen of $C/O<1$ to explain the observed features.

Altering $T_{int}$ can lead to significant effects on the thermal structure, the vertical mixing and dispersion of heavier elements. The combination of thermal dissociation, thermal inversion and different $T_{int}$ might have further implications to spectral features.

Several UHJs, like KELT-9b, HAT-P-7b WASP-76b, WASP-18b and WASP-121b, are being intensively observed in recent years and future missions are on the horizon  \citep[e.g.][]{Christiansen2010ApJ...710...97C, Doyon2012SPIE.8442E..2RD, Maxted2013MNRAS.428.2645M, Quirrenbach2014SPIE.9147E..1FQ, West2016A&A...585A.126W, Wong2016ApJ...823..122W, Mansfield2018AJ....156...10M, arcangeli2019,  Mansfield2020ApJ...888L..15M, Fu2021AJ....162..108F, Gandhi2024MNRAS.530.2885G, Landman2021A&A...656A.119L, Brogi2023AJ....165...91B, Coulombe2023Natur.620..292C, Mansfield2023Ap&SS.368...24M, Maxted2013MNRAS.428.2645M}. Among the most observed UHJs is WASP-121b which makes this exoplanet an ideal object to study UHJs in general. The WASP-121b with an equilibrium temperature of $T_{eq}$ $\sim 2360$ K orbits a F type star every $30.6$ h \citep{Delrez2016MNRAS, Mikal-Evans2022NatAs}. Recently, \citet{Mikal-Evans2023ApJ...943L..17M} published a phase curve of the UHJ made with the Near-Infrared Spectrograph on the James Webb Space Telescope (JWST hereafter). Furthermore, WASP-121b is on the schedule of JWST to measure the full orbit phase curve with the NIRISS/SOSS mode (Near Infrared Imager and Slitless Spectrograph / Single Object Slitless Spectroscopy) for cycle 1 \citep{lafreniere2022niriss2017jwst.prop.1201L} and and for cycle 2 \citep{Lafreniere2022jwst.prop.2759L}. The JWST observations from space and ground based telescopes line up with observations in the  available optical and near-IR data \citep{Evans2016detection, Evans2018AJ....156..283E_Trans, Mikal-Evans2020MNRAS, Wilson2021MNRAS.503.4787W, Ouyang2023RAA....23f5010O}. Additionally, \citet{Bourrier2020A&A...637A..36B}, \citet{Daylan2021AJ....161..131D} and \citet{Mikal-Evans2022NatAs} created photometric phase curves of WASP-121b with the TESS (Transiting Exoplanet Survey) respectively full phase orbits with WFC3 on the HST  (Wide Field Camera 3 on the Hubble Space Telescope). On the ground, several missions observed WASP-121b at high resolution \citep{Gibson2020MNRAS.493.2215G,Hoeijmakers2020A&A...641A.123H, merritt2020, Merritt2021MNRAS.506.3853M}. All these observations turn WASP-121b into one of most observed UHJ across wavelengths and at high resolutions in recent years.
Moreover, \citet{Parmentier2018A&A...617A.110P} and \citet{Mikal-Evans2022NatAs} run simulations for WASP-121b with SPARC/MITgcm 3D GCM.

In this study, we aim to isolate the effect of internal temperature on the mixing and transport of cloud particles in the atmospheres of UHJs. In order to isolate the effects of the internal temperature, we aim for a simple model. We couple mixing length theory (MLT hereafter) to the Exo-FMS GCM to estimate the extent and strength of convective motions in the atmosphere in order to test if convective motions can play a role in shaping the cloud structure on these exoplanets.
We perform a similar study to \citet{parmentier2016} but include a $K_{zz}$ diffusive mixing component to mimic mixing by convection. For simulating clouds, we use tracer based equilibrium cloud model in \citet{Komacek2022PatchyClouds}. Regarding the chemical species, we simulate only $Al_2O_3$ clouds for simplicity. For the parameterisation, we simulate an idealised planet close to WASP-121b (see Table \ref{tab:sim_paras}).

Section \ref{ch:methods_UHJ_Tint} describes the setup of the GCM simulations, the mixing length theory, the cloud formation model and the radiative transfer method briefly. In Section \ref{ch:results_UHJ_Tint}, we present the GCM outputs and the post-processing. The GCM outputs are made up of the T-p profiles, $K_{zz}$-p profiles, the stream-function in tidally-locked coordinates, the heat flow, the Miles' stability condition, the vertical turbulent heat flux, volume mixing fraction of the vapour, equilibrium saturation and of the condensates, temperature maps, $K_{zz}$ maps and cloud maps. In Section \ref{ch:discussion_UHJ_Tint}, we discuss the results and compare our findings to other studies. We summarise our key findings in Section \ref{ch:conclusions_UHJ_Tint}.

\section{Methods}
\label{ch:methods_UHJ_Tint}

\begin{table*}
	\centering
	\caption{Defined parameters for the all GCM simulations. We use a C32 resolution grid ($\approx$128$\times$64 longitude$\times$latitude).}
	\label{tab:sim_paras}
	\begin{tabular}{lcccll} 
		\hline \hline
		Symbol & Value & Units & Description & Reference\\
		\hline \hline
		$R_p$ & $1.33\cdot10^{8}$ & \lbrack m\rbrack & Planet radius & \citet{lee2022}\\
		$g$ & $10$ & \lbrack m s\textsuperscript{-2}\rbrack& Surface gravity & -\\
		$\Omega$ & $5.7\cdot10^{-5}$ & \lbrack rad s\textsuperscript{-1}\rbrack & Rotation rate &\\
		$R_d$ & $3714$ & \lbrack J K\textsuperscript{-1} kg\textsuperscript{-1}\rbrack & Specific gas constant & -\\
		$C_p$ & $1.3 \cdot10^{4}$ & \lbrack J K\textsuperscript{-1} kg\textsuperscript{-1}\rbrack & Specific heat capacity & \citet{lee2022}\\
  		$\kappa$ & $R_d/C_p$ & - & Adiabatic coefficient & -\\
		$p_{0}$ & $1\cdot 10^{8}$ & \lbrack Pa \rbrack & Reference surface pressure& -\\
		$T_{int}$ & $100$, $200$, $300$, $400$, $500$ & \lbrack K\rbrack & Internal temperature & -\\
        $T_{irr}$ & $3000$ &\lbrack K\rbrack & Irradiation temperature & -\\
		$M/H$ & $0$ & - & $\log10$ solar metallicity & -\\
        $q_{v, \:deep}$ &$1.346\cdot 10^{-6}$ & - & Deep vapour volume mixing ratio & \citet{Woitke2018}\\
        $p_{deep}$ &$1\cdot 10^{8}$ & - & pressure for the deep  mixing & -\\
        $\alpha$  & $1$ & $-$  & MLT scale parameter & \citet{lee2022}\\
        $\beta$  & $2.2$ & $-$  & Overshooting parameter & \citet{lee2022}\\
        $K_{zz,min}$  & $10$ & \lbrack m\textsuperscript{2} s\textsuperscript{-1}\rbrack&  Minimum convective diffusion coefficient & -\\
        $\Delta t_{hyd}$ & $20$ & \lbrack s\rbrack  & Hydrodynamic time step & -\\
        $\Delta t_{rad}$ & $20$ & \lbrack s\rbrack  & Radiative time step & -\\
        $\Delta t_{MLT}$ & $0.5$ & \lbrack s\rbrack  & MLT time step & \citet{lee2022}\\
        $\sigma$ & $2.0$ & $-$  & Log-normal particle size distribution & \citet{Ackerman2001}\\
        $\tau_c$  & $10$ & $[s]$  & Condensation timescale & -\\
        $\tau_{bot,initial}$  & $1\cdot10^{4}$ & \lbrack s\rbrack  & Initial linear basal drag & -\\
        $\tau_{bot}$  & $1\cdot10^{5}$ & \lbrack s\rbrack  & Rayleigh dampening timescale & -\\
        $p_{\tau,bot}$  & $1\cdot10^{7}$ & \lbrack Pa\rbrack  & Rayleigh pressure threshold& -\\
        $N_v$  & $54$ & $-$  & Vertical resolution & -\\
        $d_4$  & $0.16$ & $-$  & $\mathcal{O}(4)$ Divergence dampening coef. &\citet{lee2022} \\
		$t_{tot}$ & $2100$ & \lbrack Earth days\rbrack & Run length & -\\
  
		\hline
	\end{tabular}
\end{table*}

For this study, we use the Exo-FMS GCM \citep[e.g.][]{lee2021} with the following physics modules:
\begin{itemize}
  \item mixing length theory \citep[e.g.][]{Lee2024MNRAS.529.2686L},
  \item vertical diffusive tracer mixing \citep[e.g.][]{Lee2024MNRAS.529.2686L},
  \item tracer based equilibrium cloud model \citep{Komacek2022PatchyClouds},
  \item picket-fence opacity radiative-transfer scheme approach \citep{lee2021}.
\end{itemize}
For the initial T-p conditions, we use the picket-fence analytical T-p profile solution from \citet{parmentier2015}.
For simplicity, we do not include the effects of hydrogen dissociation and recombination in our simulations, which can have effects on the dynamical structure of the UHJs exoplanets \citep[e.g.][]{Tan&Komacek2019}.
We run the simulation for 2000 days. Afterwards, continued for 100 days to take the average as the final result. To stabilize the simulations in the deep atmosphere, we include a linear Rayleigh ‘basal’ drag similar as in \citet{Tan&Komacek2019} and \cite{Lee2024MNRAS.529.2686L}, specified in Table \ref{tab:sim_paras}.
Below, we briefly summarise each of the physics modules used in the current study.

\subsection{Mixing length theory}
\label{ch:mixing_length_theory}

Global scale GCMs generally cannot resolve convective motions which occur at scales of kilometers and less. We approximate the sub-grid processes of convection by including mixing length theory (MLT hereafter) inside the Exo-FMS model. We follow the simple MLT approach from \citet{Joyce2023}. The MLT adjusts the vertical temperature structure by an convective temperature tendency \citep[see Equation 7 in][]{Lee2024MNRAS.529.2686L}. The temperature tendencies on the layers depends on the vertical convective heat fluxes \citep[see Equation 3 in][]{Lee2024MNRAS.529.2686L} from the levels. The temperature gets changed where the local lapse rate (vertical temperature gradient) exceeds the adiabatic lapse rate ($\nabla > \nabla_{ad}$).

Moreover, MLT can estimate the convective (thermal eddy) diffusion coefficient, $K_{zz}$ \lbrack m\textsuperscript{2}s \textsuperscript{-1}\rbrack, through the relation \citep{Marley2015} 
\begin{equation}
\label{eq:thermal_eddy_diffusion_coefficient }
K_{zz} = wL,
\end{equation}
where $w$ \lbrack m s \textsuperscript{-1}\rbrack \:is the characteristic vertical velocity and $L$ \lbrack L\rbrack \:the characteristic mixing length. Additionally, we parameterise overshooting of convective motions following \citet{Woitke2004}.
All tracers diffuse horizontally and vertically through the advective component, and additionally vertically through the new convective component, the eddy diffusion coefficient like in \citet{Lee2024MNRAS.529.2686L}. Similarly, we follow the first order explicit time-stepping method like in \citet{Tsai2017} to compute the vertical diffusion of tracers inside the GCM.

\subsection{Cloud formation model}
\label{ch:cloud_formation_model}

We use a simple tracer based equilibrium cloud formation model based on \citet{TanShowman2021a}, \citet{TanShowman2021b}, and \citet{Komacek2022PatchyClouds} coupled to Exo-FMS. This scheme uses a relaxation timescale method, converting the condensable vapour volume mixing ratio, $q_v$ to the condensed vapour volume mixing fraction, $q_c$, and vice-versa depending on the equilibrium saturation volume mixing ratio, $q_s$ and parameterised condensation timescale $\tau_{c}$ [s]. 
Following the results of \citet{parmentier2018}, we assume an Al$_{2}$O$_{3}$ cloud particle composition with a log-normal size distribution with a median particle size of $r_m =1$ {\textmu}m and standard deviation of $\sigma$ $=$ $2$, and condensation timescale of $\tau_c= 10$ s.

\subsection{Picket-fence radiative-transfer}
\label{ch:picket_fence_RT}

For the radiative transfer, we use the non-grey "picket-fence" scheme from \citet{lee2021}. The picket-fence approach simulates the radiation propagating in 3 short-wave visible and 2 long-wave infrared bands vertically through the atmospheric layers. For each long-wave band, the scheme uses two representative opacities: the molecular and atomic line opacity and the general continuum opacity derived from the Rosseland mean opacity \citep{parmentier2014, parmentier2015}. In this study, we ignore any effects of cloud opacity or radiative feedback from clouds inside the GCM. But we include radiative clouds later in the post-processing (see Section \ref{ch:post-processing_UHJ_Tint}).

\section{Results}
\label{ch:results_UHJ_Tint}

In this section, we detail the results of the GCM simulations with different $T_{int}$. Then, we post-process the GCM output to produce synthetic spectra of each of the simulations.

\subsection{GCM outputs}

\begin{figure*}[t!]
	\includegraphics[width=0.95\textwidth]{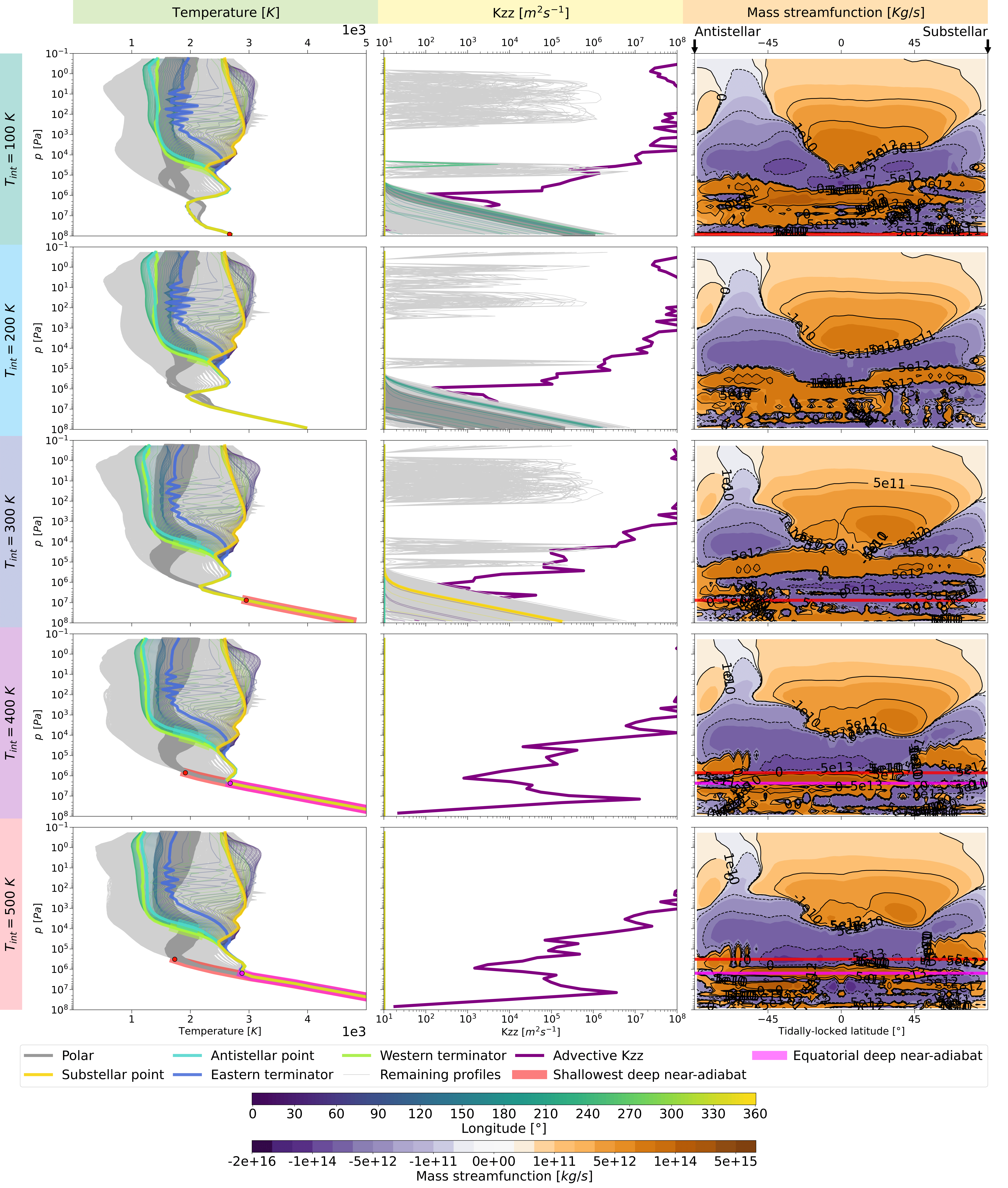}
    \caption{T-p profiles (first column), convective and advective $K_{zz}$ profiles (second column) and overturning circulation depicted by the stream-function $\Psi '$ (third column) for each simulation. The coloured lines indicate vertical profiles along the equator and its coordinates by the colourbar. The dark grey lines shows T-p profiles at the latitudes 87°N and 87°S. The bold coloured lines represent profiles at the western, eastern terminators, sub- and antistellar points. The brighter grey lines represents all other vertical profiles. The shaded area indicates the adiabatic regions. The red and magenta shaded area represent the shallowest adiabatic RCB profiles respectively along the equator. The points in column 1 and the horizontal lines in column 3 show the lowest pressure of the near-adiabatic lines. The purple line in column 2 shows the absolute advective $|K_{zz}|$ according Equation \ref{eq:Kzz_chamberlain}. The stream-function shows anticlockwise and clockwise circulations in the orange in purple colours respectively.}
    \label{fig:temp_Kzz_diff_profiles}
\end{figure*}

Figure \ref{fig:temp_Kzz_diff_profiles} presents the T-p profiles, convective and advective $K_{zz}$ profiles, and the relative difference in the T-p profiles compared to the $T_{int}$ $= 500$ K simulation. Looking at the temperature structure first, the temperature follows the near adiabatic gradient at higher pressures (below the RCB) at higher $T_{int}$. An increase of $T_{int}$ leads to lower pressure of the RCB. Comparing differences between the T-p profiles, temperatures deviate significantly in the upper atmosphere ($p<10^{4}$ Pa) on the nightside like at the antistellar point, western and eastern terminator between each simulation, showing how the variation of $T_{int}$ affects the photospheric temperatures above the RCB.
Our results show a temperature variation of between $200$ and $500$ K due to the difference in $T_{int}$ in the upper atmosphere, which suggest $T_{int}$ has significant effect on the temperature and dynamical structure of the atmosphere. These results are in line with the results of \citet{Powell2018}, when they compared T-p profiles with different entropy.

Before looking at convective regions, it is worth looking at regions where the T-p profiles are near-adiabatic conditions (with 5 \%). There are shallow near-adiabatic regions around the antistellar point and the western terminator at pressures $pa < 10^5$ Pa. These near-adiabatic regions lie on the lower end of the superrotating jet. At higher pressures, we find deep atmospheric layers with near-adiabatic conditions. We determined the RCB around the upper end of these near-adiabatic layers (following \citet{thorngren2019}). The RCB gets shallower the higher the $T_{int}$ is set. Moreover, the pressure of the RCB significantly varies across the globe. We find the lowest pressure of the RCB around the mid and high latitudes. Whereas the RCB moves to higher pressures around the poles and along the equator. But the RCB is nearly homogeneous along the equator. Lower $T_{int}$ does lead aside the sinking of the RCB as well to much lower temperatures in the deep atmosphere ($p > 10^6$ Pa). The lower temperatures occur in a region where the temperature structure, convective activities and overturning circulation differ significantly between the cases. Higher convective activities, different advection and inhomogeneities in the opacities due to different temperature structures can lead to a significant cooling in the deep atmosphere which we will explore hereafter.

Looking at the $K_{zz}$ profiles, the $T_{int}$ $= 400$ K and $500$ K simulations are fully statically stable, while variations in the convective $K_{zz}$ occur in simulations with $T_{int}$ $\leq 300$ K. Furthermore, the convective eddy diffusion coefficient generally remains low in each simulations at the imposed minimum value. The global advective $K_{zz}$ (shown in purple) in the 3 colder cases surpasses the convective $K_{zz}$ throughout the most of the atmosphere, except for pressures $10^4\lesssim p \lesssim 10^5$ Pa and in the deep atmosphere $p \gtrsim 10^6$ Pa. The convective $K_{zz}$ remains negligible in the entire atmosphere in the 2 warmest cases. Only the colder cases show a few regions with stronger convective mixing below and above the RCB. The low convective regions coincide with low lapse rates above the RCB due to the strong irradiation. Per se, convection is inhibited when the local lapse rate is much higher than the adiabatic lapse rate. The absolute stable condition is mostly met in the atmosphere above the RCB. Below the RCB, lower $T_{int}$ lead to higher convective mixing than above the RCB in the colder cases. Disturbances like waves and warming effects around the lower boundary may trigger convection in these temporarily less stable layers.

\begin{figure}[h!]
	\includegraphics[width=0.95\linewidth,left]{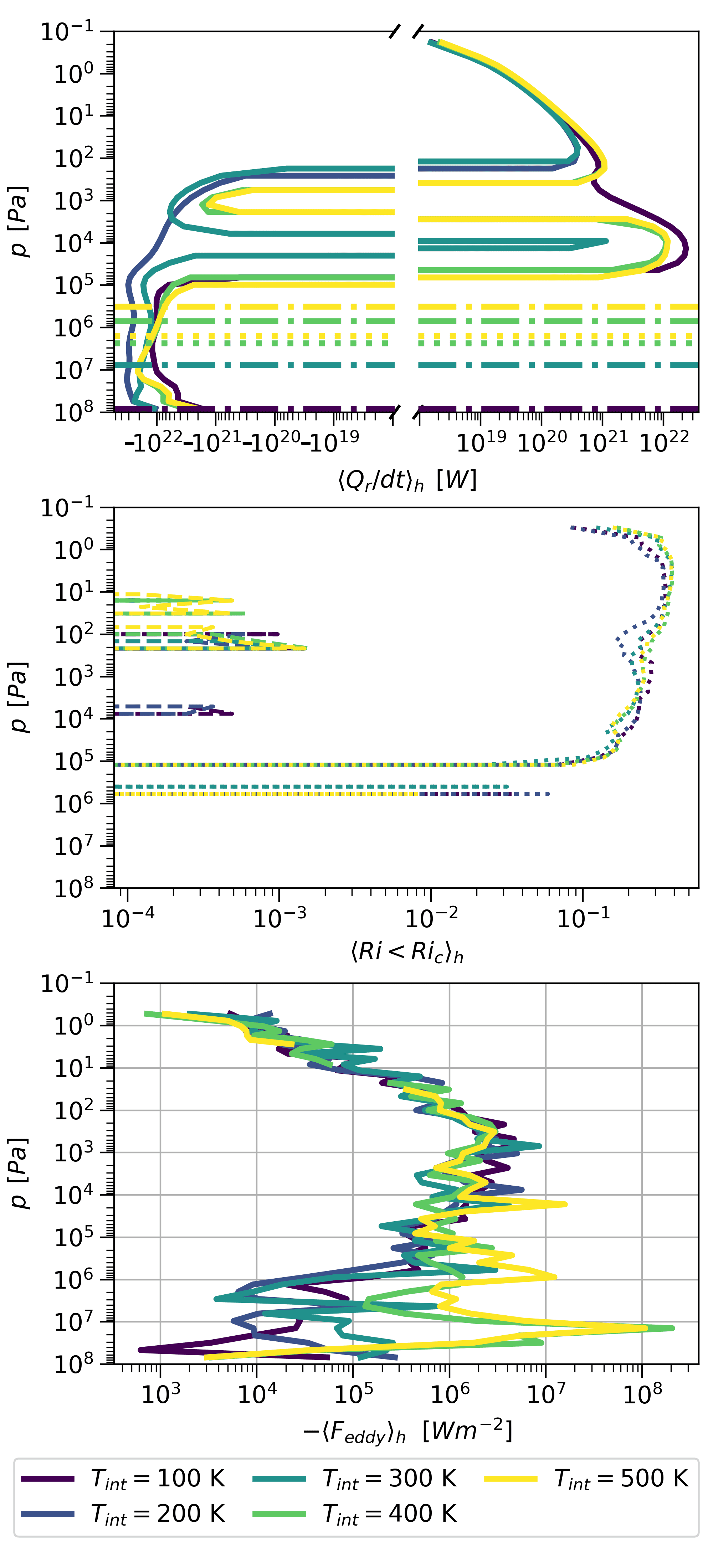}
    \caption{
    The vertical heat flow $\langle Q_r/dt \rangle_h$ \:in the first row (upwards heat flow is positive), horizontally averaged Miles’ instability condition $\langle Ri<Ri_c \rangle_h$ \:profiles in the second row, and the horizontally averaged absolute vertical turbulent heat flux $-\langle F_{eddy}\rangle_h$ \:in the third row. The coloured lines indicate the simulations with different $T_{int}$. The horizontal lines in the top figure show the lowest pressure of the near-adiabatic lines. The dotted and dash-dotted lines in the top figure indicate the equatorial RCBs respectively the shallowest RCB for the related $T_{int}$. The solid, dashed and dotted lines indicate $Ri_c$ with 0.25,  1 and 10. The dashed and solid lines indicate the $-\langle F_{eddy}\rangle_h$ \:calculated with the advective respectively the total $K_{zz}$.
    }
    \label{fig:advection_Heat_and_Ri}
\end{figure}

Why convection is inhibited in the warmer cases can be answered by the overturning circulation depicted by the stream-function $\Psi '$ in column \hyperref[fig:temp_Kzz_diff_profiles]{3}. The stream-functions $\Psi '$ of the 5 simulations contain 2 major overturning cells in the upper atmosphere and several smaller cells in the deep atmosphere. The top major cells transports gas and heat from pressures $p<10^{4}$ Pa ($p<10^{5}$ Pa for $T_{int}= 100$ K) upwards on the dayside and then at lower pressures to the nightside. The next lower major circulation cell transports gas on the dayside downwards up to pressures between $p \sim 10^{4}$ and $\sim 10^{6}$ Pa. The lower major cell is larger at the lower extend in the case with $T_{int}= 400$ and $500$ K. The case with $T_{int}= 300$ K an additional circulation cell within the space where the other cases evolved the lower major cell. The net vertical heat flow by these circulation cells is illustrated in Figure \ref{fig:advection_Heat_and_Ri}.

Figure \ref{fig:advection_Heat_and_Ri} shows the vertical heat flow $\langle Q_r/dt \rangle_h$, the Miles’ instability condition $\langle Ri<Ri_c \rangle_h$, and the turbulent heat flux $-\langle F_{eddy}\rangle_h$. The horizontally averaged vertical heat flow is directed upwards for all cases at pressures $p \lesssim 10^2$ Pa. This upward heat transport corresponds to the upper overturning circulation cell. At pressures around $p \sim 10^3$ Pa, the overall circulation transports heat downwards in all cases except for the coldest case. The change in the heat transport corresponds the transition from the upper to the lower overturning circulation cell. The coldest case has the vertically largest and the strongest upper circulation cell. Therefore, the coldest case has a net upward heat flow in the upper atmosphere, at pressures $p \lesssim 10^5$ Pa. The 3 warmest cases have evolved a net upward heat flow in the upper half of the lower overturning circulation cell, around $p \sim 10^4$ Pa. The second coldest case has a lower circulation cell which does not transport heat so efficiently upwards as it does downwards. In this way, the second coldest case shows a net downward heat transport from low pressures to the lower boundary ($p \gtrsim 10^2$ Pa). In the deeper atmosphere ($p \gtrsim 10^5$ Pa), all cases have evolved a net downward heat flow. The downward transport of gas by the lower circulation cells implies the flow of high potential temperature fluid parcels from the lower photosphere deeper into the atmosphere (below the RCB in the hotter cases). This advection of potential temperature contribute to the inhibition of convection, especially for the warmer cases.

Another way convection gets inhibited, is the presence of turbulences and the driven heat flux. The Miles’ instability condition can show if a fluid flow is turbulent. The horizontally averaged Miles’ instability condition in row \hyperref[fig:advection_Heat_and_Ri]{2} is only significantly met at pressures $p \lesssim 10^5$ Pa if we set $Ri_c = 10$. There is a less significant peak at pressure just below $10^6$ Pa with $Ri_c$ of 10 and 1 where lower $T_{int}$ yields higher values. Even fewer air masses with met Miles’ instability condition of $Ri_c= 1$ are met at pressures $p \lesssim 10^4$  Pa. Therefore, turbulences may occur with small likelihood at pressures $p \lesssim 10^6$ Pa. But at pressures $p > 10^6$ Pa, there is not a clear sign of turbulences. But it does not exclude the presence of turbulences, since they can exist up to $Ri_c= 10$ and beyond \citep{Ostrovsky2024NPGeo..31..219O}.

Additional to the Miles’ instability condition, we look at the turbulent heat flux. The horizontally averaged vertical turbulent heat flux $-\langle F_{eddy}\rangle_h$ \:in row \hyperref[fig:advection_Heat_and_Ri]{3} increases from low pressures up to $p \sim 10^3$ \: Pa for all $T_{int}$. Deeper into the atmosphere, the $-\langle F_{eddy}\rangle_h$ of 3 coldest cases decrease in waves with higher pressures. The 2 warmest cases decrease significantly less than the colder cases, but the peaks in the wavy pattern increase with depth. The much higher $-\langle F_{eddy}\rangle_h$ of 2 warmest cases can inhibit contribute enough energy to inhibit convection in the deep atmosphere.

The lowering of $T_{int}$ leads to lower $-\langle F_{eddy}\rangle_h$ and to higher convection and cooling in the deep atmosphere. The lower $T_{int}$ as well evolves different overturning circulation cells and heat flow patterns. The differences in temperature structures can get even amplified by inhomogeneities in the opacities and thereby altered fluxes. In combination, these 3 factors lead to a significant cooling in the deep atmosphere when $T_{int}$ gets lowered. We will examine the consequences in the temperature structure and advection by looking at the vapour and cloud tracers.

\begin{figure*}[t!]
	\includegraphics[width=\textwidth]{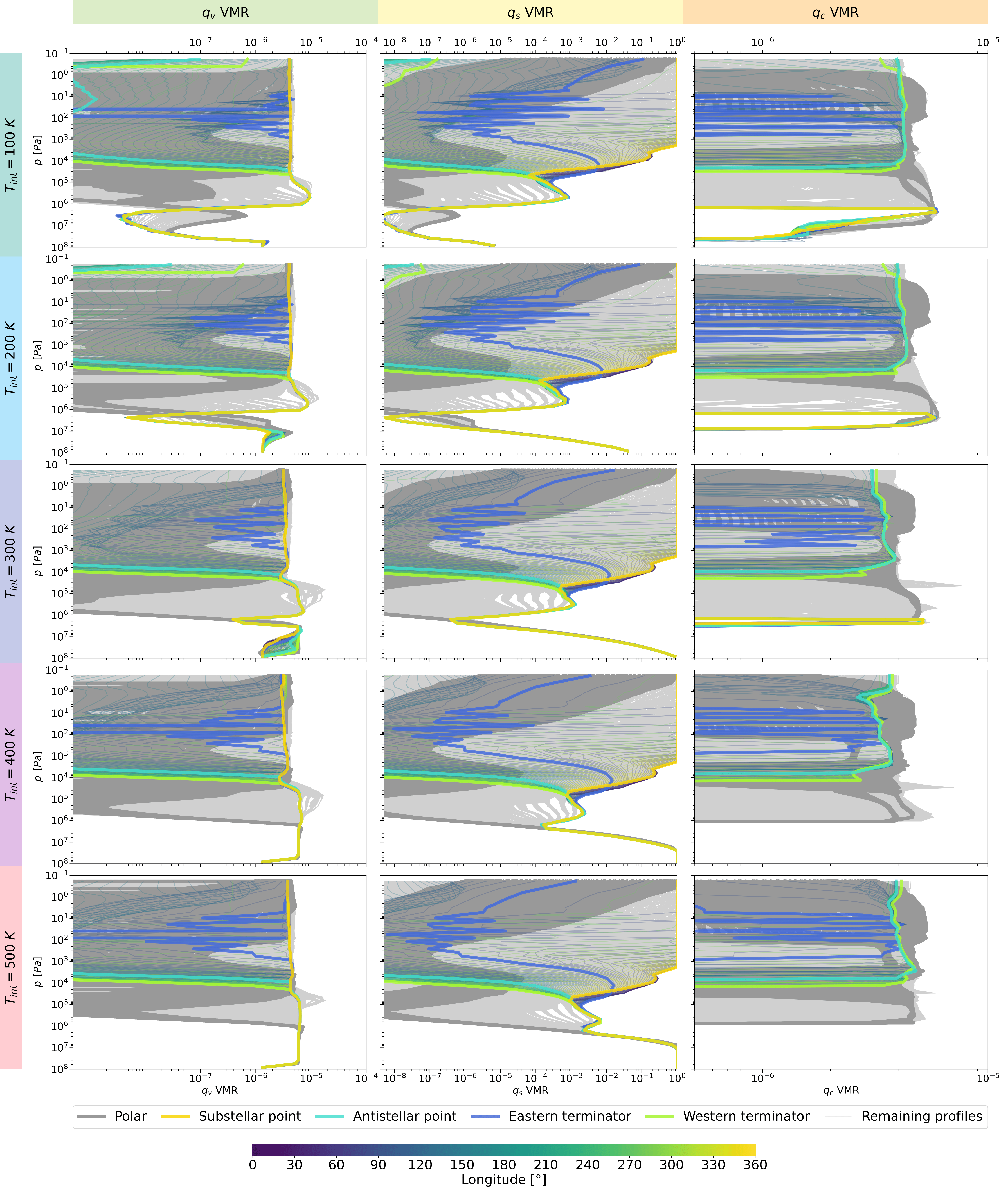}
    \caption{$q_v$ (first column), $q_s$ (second column) and $q_{c}$ profiles (third column) for each simulation. The coloured lines indicate T-p profiles along the equator and its coordinates by the colourbar. The dark grey lines show T-p profiles at the latitudes 87°N and 87°S. The bold coloured lines represent T-p profiles at the western, eastern terminators, sub- and antistellar points. The brighter grey lines represents all other T-p profiles.}
    \label{fig:q_v_q_s_q_c_profiles}
\end{figure*}

Figure \ref{fig:q_v_q_s_q_c_profiles} shows profiles of the the condensable vapour volume mixing ratio, $q_v$, the equilibrium saturation volume mixing ratio, $q_s$, and the condensed vapour volume mixing fraction,$q_c$,  for each simulation.
For the vapour, the region around the substellar point contains the most vapour above the RCB where cloud particles are vaporised on the dayside. Around the eastern terminator the vapour fraction lowers due to cooling air parcels forming condensates on the way to nightside. Further downstream in the superrotating jet, the cooling progresses and condensates more vapour until reentering the dayside. Right below the RCB, the temperature inversion limits the amount of vapour which is more pronounced in the colder cases. The coldest cases show a significant higher vapour fraction in these layers around the polar regions. The polar regions are warmer in the colder cases which come along with higher convective activity. At higher pressures, the vapour amount increases immensely with higher temperatures in the coldest cases. Contrarily, the warmer cases ($T_{int}\geq$ $400$ K) do not get affected by the significantly weaker temperature inversion. Thereby, the vapour amount remains high below the RCB, except for the lower boundary. There, the deep vapour volume mixing ratio, $q_{v, \:deep}$, constrains the vapour fraction, except for the coldest case. The constraint leads to higher vapour fraction in the layers above in the warmer cases, since $q_s$ is higher than $q_v$ or $1$.

Comparing the cloud structures, the vertical extent of the clouds increases the lower the $T_{int}$ is set. Therefore, the cloud base lowers due to lower $q_s$ responding to lower set $T_{int}$. In the superrotating jet, clouds start to form while entering cooling regions in the east with lowering $q_s$. These clouds get more thick as cooling continues and as $q_s$ lowers. Reentering the dayside, the higher irradiation warms up the gas and increases the $q_s$. Therefore, the higher $q_s$ leads to the evaporation of the clouds. In general, the cloud thickness around the poles mostly surpasses those around the equator in all cases. Additionally, the cases with $T_{int}= 300$ and with $400$ K predict less dense clouds along the equator than the other cases in the upper atmosphere (pressures $p<10^4$ Pa). This trend gets stronger below the photospheric zone (at pressures $10^4<p<10^6$ Pa). At higher pressures ($p>10^6$ Pa), low level clouds cover the entire planet at lower $T_{int}$. 

Figure \ref{fig:temp_Kzz_diff_profiles} shows the convective $K_{zz}$ to be generally low in the atmosphere compared to the advective mixing. However, clouds are ubiquitous in every simulation. This suggest that convection play a very minor role in setting the cloud coverage in UHJs compared to advection.

\begin{figure*}[t!]
	\includegraphics[width=\textwidth]{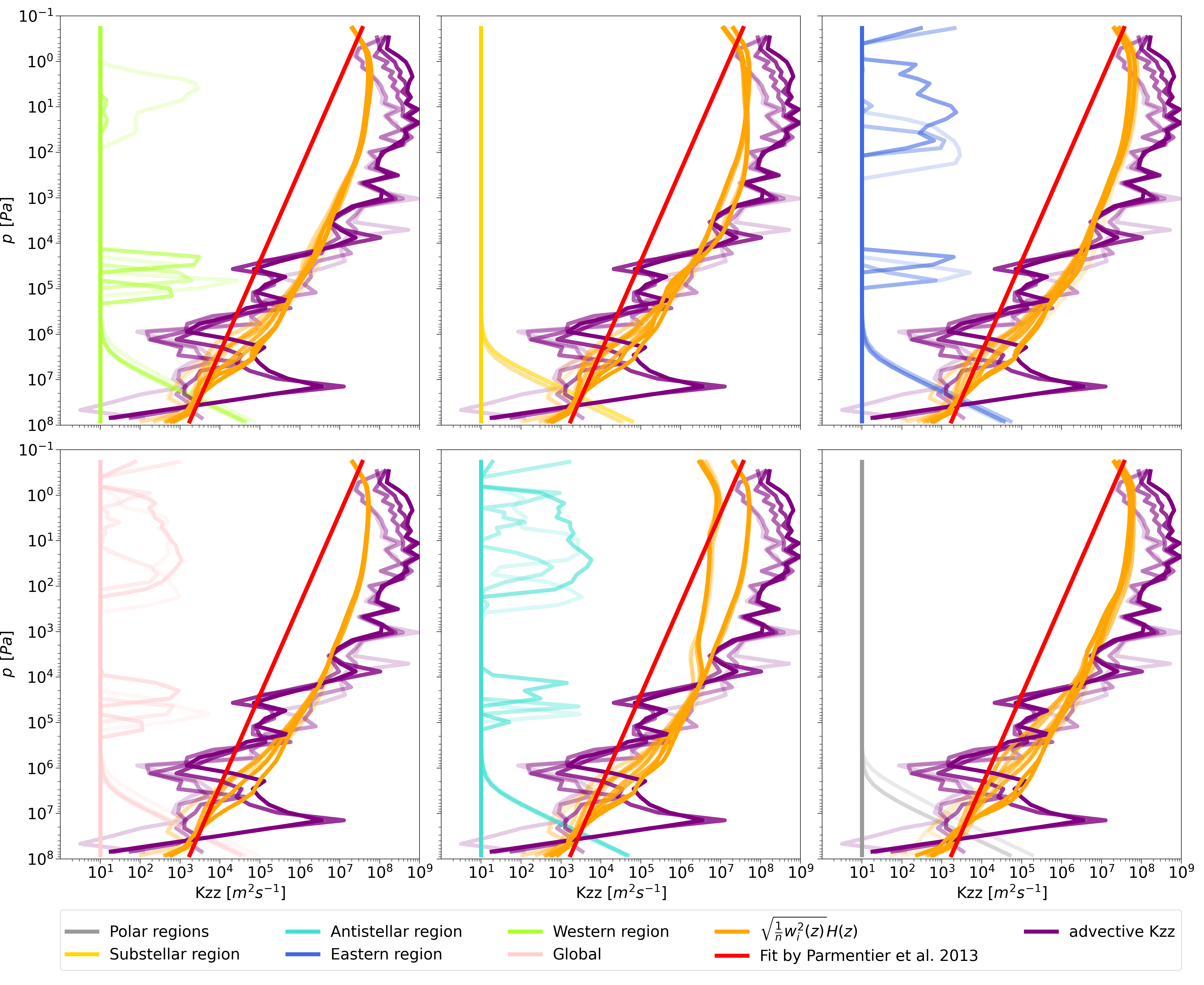}
    \caption{Average $K_{zz}$-p profiles of the western region ($x>225^\circ$ and $x<315^\circ$), substellar region ($x\geq-45^\circ$ and $x\leq45^\circ$), eastern region ($x>45^\circ$ and $x<135^\circ$), global, antistellar region ($x\geq45^\circ$ and $x\leq135^\circ$), polar region ($y \geq66.565^\circ$ or $y \leq66.565^\circ$). For comparison, the common estimate of $K_{zz}$ in the literature \citep{Lewis2010,Moses2011} as $K_{zz}=H(z)\sqrt{\langle w^2\rangle_h}$, and the 1D fit by \citet{Parmentier2013} as $K_{zz}=5\cdot10^4\sqrt{\frac{p}{10^5}}$ are shown in each subplot. All the subplots show regional averages at altered $T_{int}$. The more faint the line colour, the lower the $T_{int}$ value. The purple line shows the global absolute advective $K_{zz}$ according Equation \ref{eq:Kzz_chamberlain}.}
    \label{fig:Kzz_regional_profiles}
\end{figure*}

Figure \ref{fig:Kzz_regional_profiles} show regional averaged $K_{zz}$-profiles. Where convection appears in the upper atmosphere, $p<10^3$ Pa, it is confined to the flanks of the superrotating jet and inside the nightside Rossby gyres. Although the lapse rate is very low along the equator, the high latitudinal edges have higher lapse rates. In this regions, the layers below tend to be warmer in the colder cases. The combination with strong cooling from above and the warming effect from below leads to situations with high enough lapse rate to trigger convection. This happens in the eastern region in the colder cases. A similar mechanism applies for the antistellar and western region for the colder cases. A wider warmer jet than the warmer cases advects warmer gases to the nighside, like at pressure $p=10^3$ Pa. The warmer gases lie below colder and more low latitude gyres which leads to high enough lapse rates for convection. The cold cases vary in the combination of the temperature of the jet and the extent of the gyres. While the case with $T_{int}= 300$ K has larger gyres at low pressure, the colder cases have a warmer jet.

The deepest convective zone extends from the lower boundary to pressures around the RCB and is present only in the colder cases ($T_{int} \leq 300$ K). The colder cases show $K_{zz}$ values decreasing with height and no effect of the strong convective inhibition due to the inversion in the colder cases. The colder the deep atmosphere gets the higher the $K_{zz}$ in the polar region becomes.

\begin{figure*}[t!]
	\includegraphics[width=\textwidth]{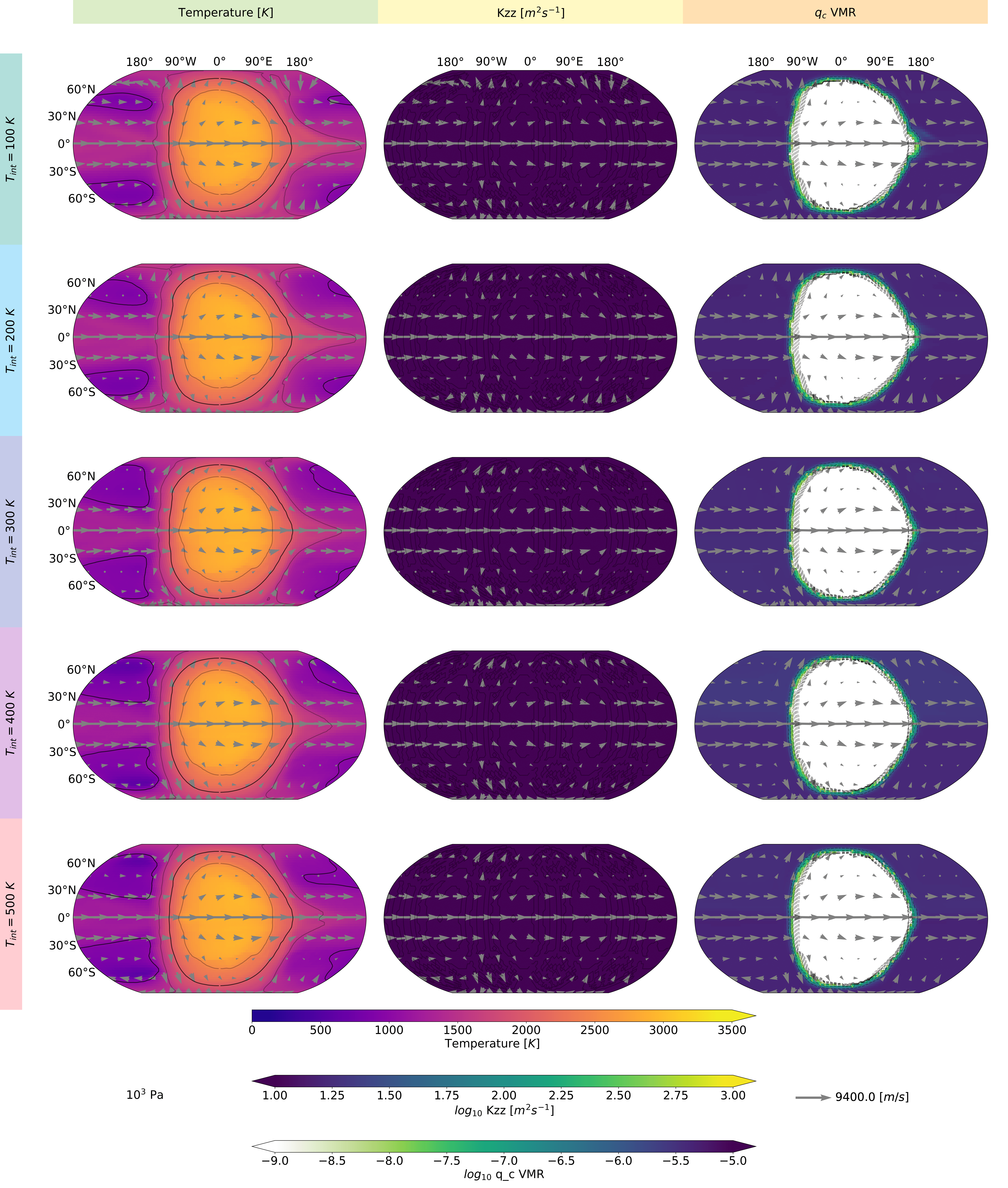}
    \caption{Maps of the temperature in column 1, convective diffusion coefficient $K_{zz}$ in column 2, and cloud tracer $q_{c}$ in column 3 at pressure surface $p=10^{4}$ Pa. Each row shows the simulation outputs with different $T_{int}$. The arrows indicate horizontal wind speed and direction.}
    \label{fig:temp_Kzz_q_c_10e4_maps}
\end{figure*}

Figure \ref{fig:temp_Kzz_q_c_10e4_maps} shows maps of the temperature, $K_{zz}$, and of $q_{c}$ at $10^{4}$ Pa with different $T_{int}$. Looking at the temperature in column \hyperref[fig:temp_Kzz_q_c_10e4_maps]{1}, the size of the hotspot increases as $T_{int}$ is set higher. Furthermore, higher $T_{int}$ leads to the advection of warmer gases along the jet which increases the temperature the higher the $T_{int}$ gets. The temperatures inside the gyres rise when $T_{int}$ is increased from $100$ to $300$ K. Then, the gyres cools again at higher set $T_{int}$. The cooler gyres come along with more cyclostrophic flow.

Looking at the spatial variation in column \hyperref[fig:temp_Kzz_q_c_10e4_maps]{3}, the cloud thickness changes the most on the edge of the hotter dayside and smoothly from the equator to higher latitudes. These few gradients and mostly uniform cloud cover suggest the advective nature of the clouds rather than the patchy pattern expected from convection. The relatively low $K_{zz}$ support this argument.

\begin{figure*}[t!]
	\includegraphics[width=\textwidth]{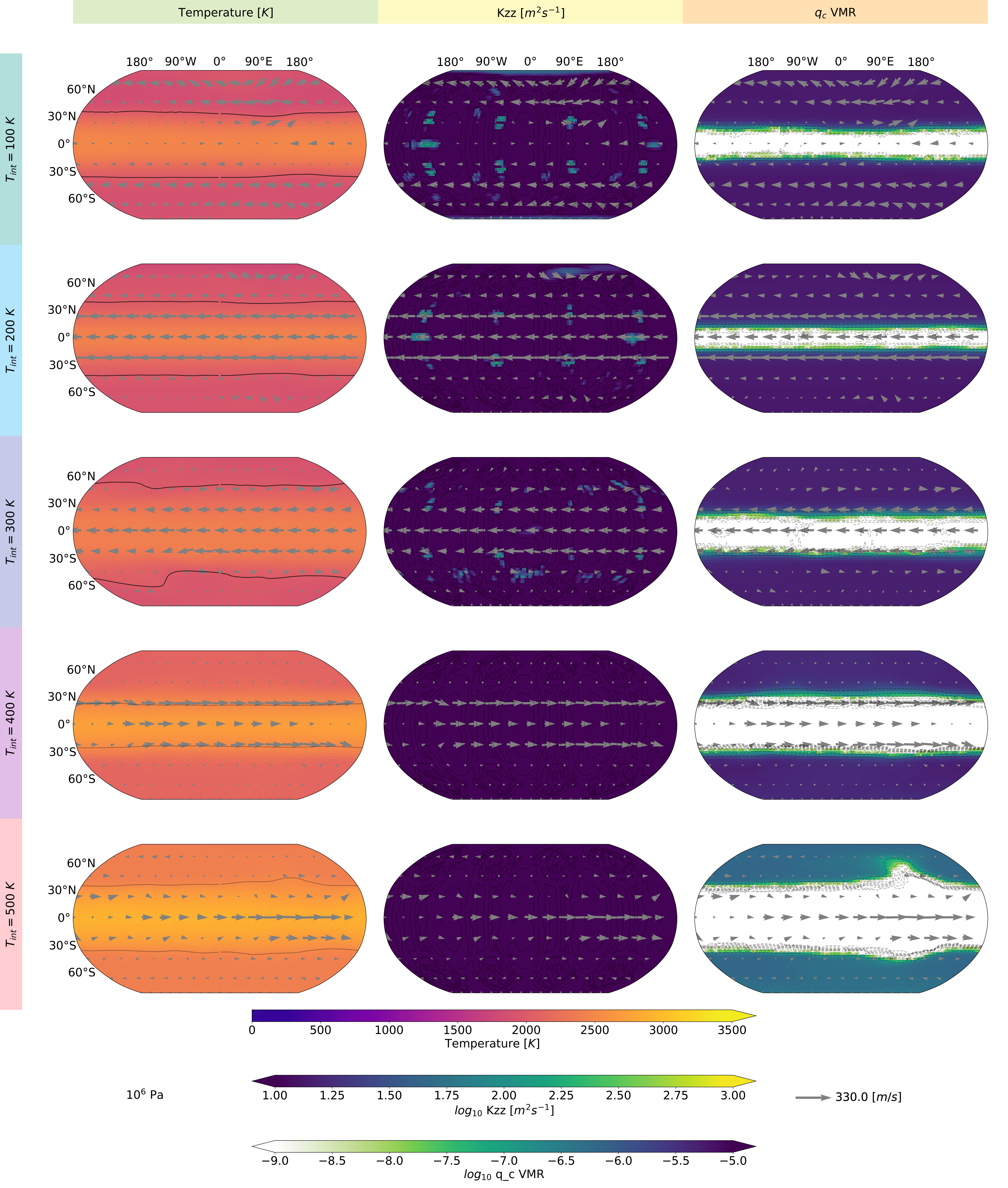}
    \caption{Maps of the temperature in column 1, convective diffusion coefficient $K_{zz}$ in column 2, and cloud tracer $q_{c}$ in column 3 at pressure surface $p=10^{6}$ Pa. Each row shows the simulation outputs with different $T_{int}$. The arrows indicate horizontal wind speed and direction.}
    \label{fig:temp_Kzz_q_c_10e6_maps}
\end{figure*}

Going to higher pressures, the figure \ref{fig:temp_Kzz_q_c_10e6_maps} shows maps of the temperature, $K_{zz}$, and of $q_{c}$ at $10^{6}$ Pa with different $T_{int}$. Here in the deep atmosphere, the temperatures respond to the set $T_{int}$ more directly (see column \hyperref[fig:temp_Kzz_q_c_10e6_maps]{1}). Moreover, the low latitudes are warmer than the higher latitudes in all cases. The warmer temperatures come along with less cloud cover over the equator (see column \hyperref[fig:temp_Kzz_q_c_10e6_maps]{3}). The highest cloud cover is reached by the case with $T_{int}=200$ K at this pressure. Except from this case, the higher the $T_{int}$ is set the smaller the $q_c$ gets around the equator. Whereas the colder cases ($T_{int}\leq300$ K) have already the deep, westward wind direction, the warmer cases evolve equatorial winds with an eastwards flow. The difference in the evolved wind patterns derives from the varied temperature structure due to altered $T_{int}$ and thereby changed feedback. In this regard, the vertical extent and magnitude of the inversions lower the higher $T_{int}$ is set. As such, we probe different spheres with different evolved dynamics, cloud and temperature structure. These consequences in the deep atmosphere feedback with shallower layers leading to differences up there.

Although we are in convective inhibiting layers at $10^{6}$ Pa, which extends vertically and gets more intense the lower the $T_{int}$ and the lower the latitudes are, we see convective activity and mixing nevertheless (see column \hyperref[fig:temp_Kzz_q_c_10e6_maps]{2}). This higher convective activity comes from overshooting of the higher convective activity in the inversion.

Although the convective activities surpasses by far those in the upper layers, the cloud structure is similarly more or less uniform with a few gradients like at lower pressures. The cloud structure does not change noticeably if $K_{zz}$ is locally higher. Once more, the missing patchy patterns in the cloud structure show the advective nature of the clouds, even in the deep atmosphere.

\subsection{Post-Processing}
\label{ch:post-processing_UHJ_Tint}

\begin{figure*}[t!]
	\includegraphics[width=\textwidth]{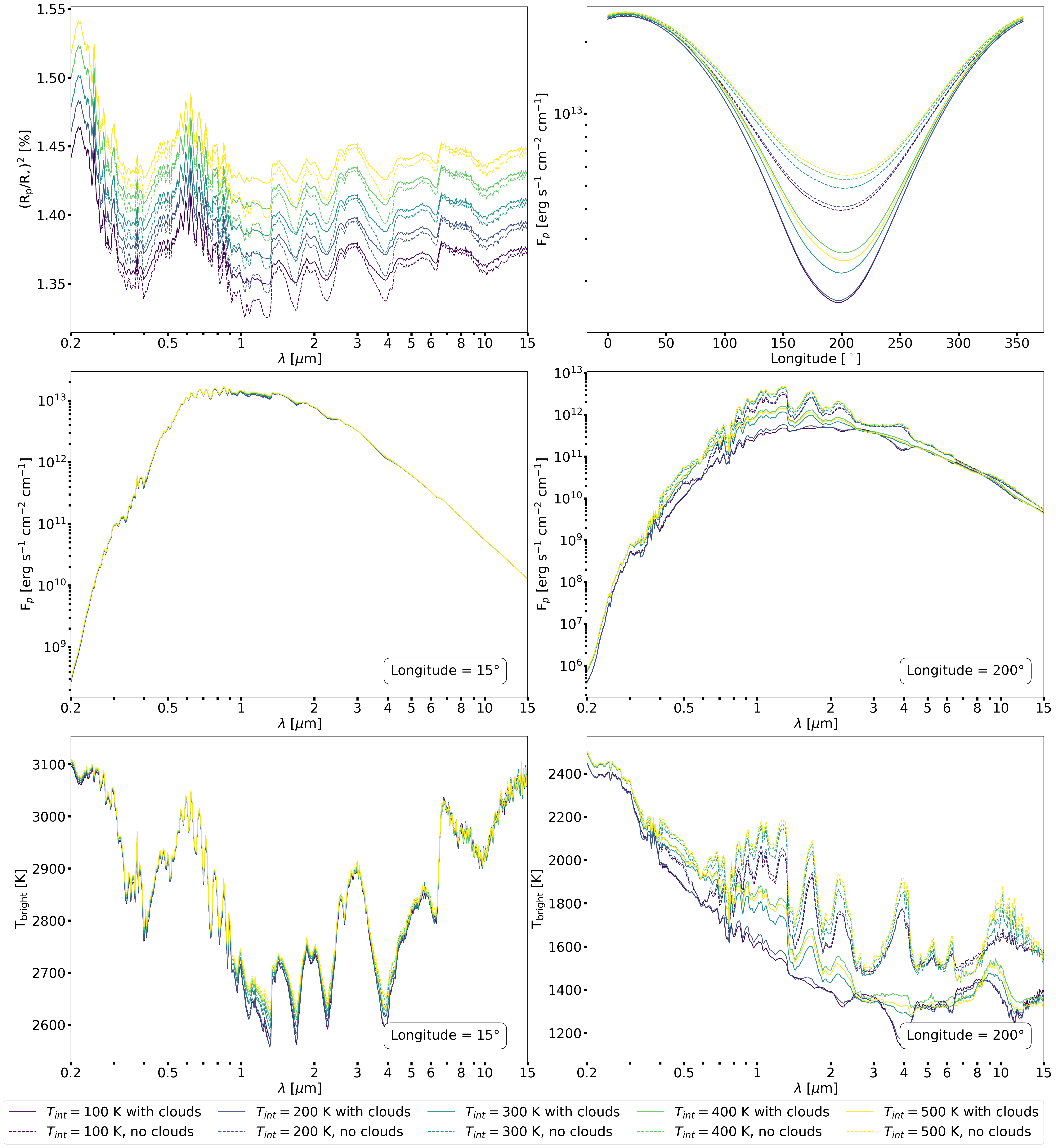}
    \caption{Transmission spectra (top left), phase curves (top right), planetary spectral flux at a longitudinal viewing angle of $15^\circ$ (center left) and $195^{\circ}$ (center right), and brightness temperature at a longitudinal viewing angle of $15^\circ$ (lower left) and $195^{\circ}$ (lower right) based on the post-processing of the GCM simulations. The line styles denote post-processing with (solid) and with no (dashed) cloud opacity.}
    \label{fig:post-processing}
\end{figure*}

In this section, we use the 3D spectral radiative transfer model gCMCRT \citep[GPU accelerated Cloudy Monte Carlo Radiative Transfer Method on a spherical geometry grid]{lee2022} to post-process the GCM results. We examine the effects of changing the internal temperature and cloud structures on the resulting transmission spectra, emission spectra and phase curves of the model output.

Figure \ref{fig:post-processing} presents the transmission spectra of each of the GCM simulations at different $T_{int}$. We post-processed the outputs of the simulation with and without the cloud opacities. Higher $T_{int}$ models lead to a larger transmission $(Rp/Rs)^2$ as previously expected. Additionally, the spectra calculated with cloud opacities lead to larger transmission $(Rp/Rs)^2$ than without cloud opacities. The smaller transmission $(Rp/Rs)^2$ calculated with cloud opacities are due to the high clouds seen on the nightside and in the dusk regions in the GCM simulation. These high clouds extinct stellar flux in layers permeable for stellar light if clouds were absent. The similar offset between all spectra calculated with and without cloud opacities suggest a similar average temperature for the cases with different $T_{int}$. The similar temperature on the dayside and the few region with varying temperatures at low pressure may lead to similar heights.

Moving on to the spectrally integrated phase curve, more consequences of the different cloud and temperature structures reveal in different offsets of the minima in the phase curve (see \hyperref[fig:post-processing]{top right} panel). The absence of cloud opacities lead to a higher emission flux due to the contributions from deeper layers, especially for the emission from the nightside where high clouds could extinct significant more flux of the planetary emission. Interestingly, the most fluxes are extincted in the second coldest case which is not expected. But the coldest case has thinner top clouds and a slightly warmer nightside. A similar feature happens in the warmest cases. The second warmest case has the least cloud cover on the nightside in the upper atmosphere (pressures $p<10^4$ Pa) based on the GCM simulation. This leads to the highest emissions from the nightside compared to the other cases although it would radiate relatively less without cloud opacities.

The planetary spectral flux from the dayside do not differ much since the emission come mostly from cloud free layers (see \hyperref[fig:post-processing]{middle left} panel). Therefore, the fluxes calculated without cloud opacities do not surpass the other fluxes at most wavelengths. Where they surpass, the fluxes come from deeper layers. The corresponding brightness temperature at these wavelengths of $\sim 2600$ K (see \hyperref[fig:post-processing]{bottom left} panel) linked with the T-p profile (see Figure \ref{fig:temp_Kzz_diff_profiles}) indicate that some fluxes come from layer with pressures $10^5$ $\lesssim$  $p$ $\lesssim10^6$ Pa. At these depth, the warmest case has the least cloud cover along the the equator. Therefore, the warmest case radiates the most at this wavelength from these layers. Nevertheless, much of these fluxes get extincted by the vapour. At many wavelengths, the coldest case emits the most on the nightside, but the least on the dayside.

On the nightside, clouds reduce significant fluxes at several wavelength ranges (see \hyperref[fig:post-processing]{middle right} panel) compared to the potential fluxes calculated without cloud opacities. The potential fluxes scale hierarchically with $T_{int}$ throughout the most wavelength ranges with absorption by cloud particles, except for 2 main wavelength ranges in the NIR. These 2 ranges corresponds to a brightness temperature of $\sim 1500$ K (see \hyperref[fig:post-processing]{bottom right} panel). This low brightness temperature suggest emissions from layers lie around the gyres where the 2 coldest cases are warmer and have top thinner clouds. Inside the gyres (at the lowest brightness temperatures), the potential fluxes are in the warmer cases due to emission from much deeper and warmer layers. Nevertheless, clouds block the emission from the lower layers and only emission from the top layers of the gyres can escape. The gyres of the coldest cases are warmer at top layers due to a stronger overturning circulation and thereby emit more. Hereby, the probed nightside thermal inversion are more pronounced in the colder cases, especially around the gyres.
Around the equator's nightside, the second warmest case shows the lowest cloud thickness in the upper atmosphere (at pressures $p<10^4$ Pa, see Figure \ref{fig:q_v_q_s_q_c_profiles}). This low cloud thickness allow emissions from the deeper layers than in other cases which we see in the effective fluxes. The corresponding brightness temperatures lie between $\sim 1200$ and $1500$ K which matches the temperature in the uppermost, equatorial layers on the nightside in the second warmest case. In the same wavelengths, the warmer cases emit mostly more than the colder cases. Moreover, warmer cases reduce the emitted fluxes not that much as the colder cases do at wavelengths with potentially higher cloud opacities. Thereby, the warmer cases smooth curve of the emission fluxes at the most wavelengths with the most extinction by clouds.

This effect of cloud opacities on the post-processing surpasses the effect of $T_{int}$ on the nightside. Regarding the wavelength, the effect of cloud opacities on the nightside unfolds mainly in the visible and in some ranges of the near infrared. Therefore, the brightness temperature is higher in these wavelengths.
Higher $T_{int}$ flattens the curves of the planetary spectral flux and brightness temperature for the post-processing with cloud opacities in some ranges in the NIR bands on the nightside.



\section{Discussion}
\label{ch:discussion_UHJ_Tint}

In this Section, we discuss our results in the context of other studies that investigate vertical mixing and cloud patterns in HJ and UHJ atmospheres.

\subsection{Convective inhibition and RCB}

As our warmest cases have evolved a near adiabatic region in deep atmosphere, convective activities are expected. But convection seem to be inhibited or the steady-state (warming to the deep adiabat) has not been reached yet. Inhibited convection could be caused by the advection of potential temperature and the turbulent kinetic transport of energy which we see significantly stronger in the two warmest cases. Similarly, \citet{guillot2002} and \citet{youdin2010} suggested such convective inhibition by the turbulent kinetic transport of energy (mechanical green house effect) can occur.
All our cases show a burial of heat by a net downward heat advection by atmospheric circulation and waves like in \citet{showman&guillot2002},\citet{ guillot2002}, \citet{tremblin2017}, \citet{sainsbury2019}, \citet{mendonca2020}, \citet{Sainsbury2023MNRAS.524.1316S}. We find a similar global overturning circulation pattern in the upper atmosphere as \citet{Sainsbury2023MNRAS.524.1316S} which drives upwards vertical and downwards transport on the dayside respectively on the nightside.

There are other mechanisms of transferring energy to the interior in UHJ like ohmic heating \citep{Batygin2010ApJ...714L.238B, Perna2012ApJ...751...59P, Batygin2011ApJ...738....1B, Huang2012ApJ...757...47H, Rauscher&Menou2013ApJ, Wu2013ApJ...763...13W, rogers2014}, and atmospheric thermal tides in asynchronous HJ and UHJ \citep{Arras2010ApJ...714....1A, Gu2019ApJ...887..228G}. But these mechanisms are not modelled in the current simulations.

Altering $T_{int}$ varies the overturning circulation cells and leads to different heat flow and temperature structure. These results are in line with the results in \citet{Komacek2022ApJ...941L..40KTintDynamics}. Moreover, these effects couple together with opacity inhomogeneities which alters the temperature structure further as in \citet{Zhang2023ApJ...957...20Z_I}, \citet{Zhang2023ApJ...957...21Z_II} and \citet{Zhang2023ApJ...957...22Z_III}. They showed that opacity inhomogeneity result in larger cooling of the interior and different pressures of RCB. The combination of the overturning circulation, heat flow, opacity inhomogeneities and altered $T_{int}$ lead to a large variations in the temperature structure and pressures of RCB in our cases. Consequently, the cloud structure responds to this combined effects. For instance, the pressure of the RCB varies significantly along the latitude and remains homogeneous along the equator like in \citet{Zhang2023ApJ...957...22Z_III}. Their weak drag shows shallowest RCB in the mid-latitudes like our warmest cases. As predicted by \citet{thorngren2019}, higher $T_{int}$ pushes the RCB to lower pressures in our study.

Our colder cases have not evolved deep or near adiatbatic thermal profile. Possible reasons for this lack can be due to a missing steady-state in deep atmosphere as \citet{Sainsbury2023MNRAS.524.1316S} noticed a radius inflation up to 50000 simulation days, or the RCBs of our colder cases actually lie beyond our simulated boundary pressure. The latter is reasonable since the pressure of the RCB increases exponentially the lower the $T_{int}$ is set. As the RCB lies beyond the model boundary, the energy fluxes interact with the lower boundary in the colder cases where the convection in the deep atmosphere is triggered. Additionally, the gradient of the potential temperature in the deep atmosphere gets smaller the lower the $T_{int}$ is set. Therefore, the deep layers get less stable the lower the $T_{int}$ is set and more vulnerable to disturbances like waves and warming. However, longer simulation times may change our results.

\subsection{Vertical mixing from convection vs advection}

In all our simulations, the global atmosphere generally shows very weak mixing due to convective motions. This is expected from previous theoretical studies \citep[e.g.][]{Parmentier2013}, due to strong irradiation inhibiting convection. However, our simulations show that isolated regions, in particular the nightside Rossby gyres can exhibit strong convective vertical mixing at upper and middle atmospheric regions on par with the advective component.

Modelling and theoretical studies have investigated mixing processes in the atmosphere such as \citet{Lewis2010}, \citet{Moses2011}, and \citet{Parmentier2013}. However, these studies mostly investigated the role of advective mixing, generally ignoring the sub-scale convective component. In Figure \ref{fig:Kzz_regional_profiles}, we compare commonly used expression of the advective $K_{zz}$-profile from the literature to our derived advective and convective $K_{zz}$-profile from the GCM simulation. Our simulations produce a partially similar magnitude offset between the $K_{zz}$ value derived from Equation \ref{eq:Kzz_Lewis_Moses} \citep{Moses2011,Lewis2010} and Equation \ref{eq:Kzz_Parmentier} \citep{Parmentier2013} as noted in \citet{Parmentier2013}. But in the upper atmosphere ($p \lesssim 10 ^4$ Pa), our advective $K_{zz}$ exceed the $K_{zz}$ values derived according to these studies, even up to 3 magnitudes in some regions. At higher pressures, our advective $K_{zz}$ lie partially below and above the other $K_{zz}$ from the literature. This suggests that the $K_{zz}$-profiles derived from Equation \ref{eq:Kzz_Lewis_Moses} may be over- or underestimated in Figure \ref{fig:Kzz_regional_profiles} for UHJ depending on the pressure. The 1D fit of $K_{zz}$ in \citet{Parmentier2013} shows that it is made for a specific HJ. Despite this, it is clear from Figure \ref{fig:Kzz_regional_profiles} that connvective mixing is magnitudes smaller in the atmosphere. Only very deep regions, $p>10^7$ Pa, show convective mixing that is consistently higher than advective mixing.


The MLT scheme used in our study provides an estimate of the convective strength and estimation of the eddy diffusion coefficient. Our study suggests that there is strong static stability and low convective fluxes in UHJ with warm interiors.
Whereas, for the colder interior temperatures, convective adjustment is triggered more often and stronger, suggesting less static stability and higher convective fluxes required to adjust the atmosphere compared to the hotter interiors.
Our results also suggest that warmer interior atmospheres contribute a stabilising effect on the upper atmosphere, reducing the likelihood of upper atmospheric regions undergoing convective adjustment.

Following our results, we suggest models that require mixing processes from the deep atmosphere include a convective component when constructing vertical $K_{zz}$-profiles. In the upper atmosphere, higher $K_{zz}$ may appear in isolated regions around and inside the Rossby gyres on the nightside. Rossby gyres play an important role in the non-equilibrium chemical structures of hot Jupiters, where they show strong differences in composition compared to other regions of the atmosphere such as the equatorial jet \citep{Drummond2020, Zamyatina2024MNRAS.529.1776Z}. Hence, we suggest for GCMs of hot Jupiters consider including a convective mixing scheme to capture this additional mixing component in the Rossby gyres.

\subsection{Cloud structure}


\citet{Komacek2022PatchyClouds} performed a tracer based equilibrium cloud formation model for a UHJ with $T_{irr}$ $=$ $3348.85$ K and $T_{int}$ $=$ $507$ K similar to our study. Their model used the $K_{zz}$ approach of \citet{Parmentier2013}. For the non-radiatively active cloud-case in \citet{Komacek2022PatchyClouds}, they predict an cloud base at pressure a bit smaller than $1\cdot10^5$ Pa for the equatorial nightside and another cloud base at about $1\cdot10^5$ Pa for high latitudinal regions on the nightside. These results disagree with equatorial cloud base at about $1\cdot10^4$ Pa in our simulation and with our cloud base for the high latitudes at around $1\cdot10^6$ Pa. The differences in the cloud base may rise from a different temperature structure as our simulations tend to be warmer along the equator and colder at high latitudes in the deep atmosphere. The altered temperature structure is mainly generated by the different radiative transfer scheme (double grey in their model) and the different dynamical core. Additionally, we use 
other cloud parametrisations ($\sigma =1$ compared to $2$, $r_0=\left[2,5\right]$ $\mu m$ compared to $r_m= 1$ $\mu m$), which leads to different equilibrium saturation volume mixing ratios and thereby different cloud structures. Above, their cloud tops lay at pressures around $1\cdot10^3$ Pa for the antistellar point and around $1\cdot10^2$ Pa for the western terminator. Whereas, our cloud tops lay beyond the model boundary. The higher cloud fractions around the western terminator and the antistellar region in our simulation likely come from lower local temperatures. These lower temperatures are due to lower set $T_{irr}$, different radiative transfer scheme and due to the missing chemical heating effects from hydrogen dissociation and recombination. Moreover, they predict patchy clouds throughout the vertical cloud extent which we see in our simulations only in the upper cloud layers. Similarly as in \citet{Komacek2022PatchyClouds}, we see higher emitted nightside flux in a cloud free than cloudy atmosphere over a wide range of wavelengths in the gCMCRT post-processing. But their phase curves show an offset of the hotspot whereas our phase curves have no offset. The mismatch is likely due to mentioned warmer temperatures and thereby shorter radiative timescales in our simulated upper atmospheres.


\citet{Roman2021} and \citet{Malsky2024ApJ} studied the impact of vertical mixing through the use of different parameterised vertical cloud extends. They used the RM-GCM with a double-grey, two-stream radiative transfer scheme in \citet{Roman2021} and \citet{Malsky2024ApJ}, with a picket-fence scheme also used in \citet{Malsky2024ApJ}, similar to our study. Additionally, they included radiatively active clouds which condense when the local temperature drops below the condensation temperature of each species. When condensation occurs, they set an immediate change to a condensate with a fraction of 10\% in \citet{Roman2021} and \citet{Malsky2024ApJ}, and to 100\% in \citet{Roman2021} of the condensible species. They assumed the abundance of the species to be constant and uniform throughout the atmosphere. For varying strength of the vertical mixing, they implement a pressure-dependent vertical gradient for particle sizes with larger particles at higher pressures. In order to invest different vertical cloud extends, they limit the clouds by force to a smaller vertical extend in the compact case and set no limit for their extended case. 
Comparing the vertical cloud extend, our cases with warmer interiors agrees more to the extended cases of HD 209458b in \citet{Malsky2024ApJ} than to their compact cloud cases. Both of their extended cases show clouds reaching the upper boundary at $1$ Pa and a cloud free dayside which is in line with our cases. But the dayside is less cloud free in their both cases and their double-grey scheme generates thinner clouds along the equator in the upper atmosphere. Similarly, we see slightly thinner clouds along the upper equator region in our warmer cases. The difference between their RT cases might not only rise from the varied RT schemes, but as well from their dual cloud fraction mode, 0 or 10\% condensed vapour. Comparing the cloud bases, the bases in \citet{Malsky2024ApJ} lay globally around  $2\cdot10^6$ Pa for all their cases of HD 209458b. These pressure heights of the cloud bases agree only with our cases with colder interiors. Moreover, there is a less densely cloudy respectively cloud free equator at pressures between $\sim 1\cdot10^4$ and $\sim 5\cdot10^5$ Pa in their picket-fence and compact cloud case. This pattern agrees with the cloud structure in our colder and warmer cases at similar pressure. The differences in pressures of cloud bases can rise from from the radiative active clouds, the differently set lower boundary, the multiple cloud species, and the lower temperature in the interior.
Compared to \citet{Roman2021}, the cloud cover in our study shows similarities to the cloud optical depth (up to $2.8\cdot10^4$ Pa) in the 100\%-condensed-cloud-case (to the trend in the simulations between $T_{irr}= 2500$ and $3500$ K). The densest $Al_2O_3$ clouds are at high latitudes on the nightside and the dayside is cloud free like in our simulations. The cloud free dayside is less deformed in their cases which might be due to weaker winds resulting from weaker day-nightside temperature contrast due to radiatively active clouds and different RT scheme. Equivalent similarities can be seen between the trends in their extended-clouds-cases, nucleation-limited-clouds-cases, and our cases. But we see less similarities to all their compact cases as the nightside features weaker high latitudinal clouds. As \citet{Roman2021} demonstrated the trend in the offsets of hotspots with increasing $T_{irr}$, we show the trend in the eastward offset of the coldspot with increasing $T_{int}$ (see Figure \ref{fig:post-processing}). But the modelled range of $T_{int}$ leads to no significant offset of the hotspot.



Next, we compare our results to the outcomes of the microphysical cloud scheme by \citet{Powell2018}. They find a similar trend in the increase in the vertical extent of $TiO_2$ and $MgSiO_3$ clouds in HJs when they set a deep cold trap. This cold trap leads to the formation of a second cloud population in the deep atmosphere depending on the cloud particle size. The 2 cloud populations extend vertically and get connected, when they set lower equilibrium temperatures. But at the higher equilibrium temperature, the deep cloud population gets vertically smaller and starts to disappear. We see such a deep cloud population ($1\cdot10^6$$<p<$ $1\cdot10^8$ Pa) as well in our colder cases. But microphysical processes, vertical mixing and settling change the cloud particle sizes as shown in their study. They predicted bimodal or irregular shape in the cloud particle size distribution which lead to more complex cloud structures. Such implications on the cloud particle distribution probably would change our results significantly. For instance, our colder cases may underestimate the efficiency of the deep cold trap due to inappropriate cloud particle distributions which may lead to too much and too high cloud formation in the upper atmosphere. Nonetheless, our results agree with \citet{Powell2018} that deep cold traps still allows significant cloud formation in the upper atmosphere.


\subsection{Limitations and future improvements}



Our study focus on and isolate the effect of the $T_{int}$ and mixing on the cloud structure. Therefore, we did not include radiatively active clouds as \citet{Roman2021}, \citet{Komacek2022PatchyClouds} and \citet{ Malsky2024ApJ}, although they showed clouds change the temperature structure and atmospheric dynamics. Firstly, such radiatively active clouds lead to an increase of the scattering at short wavelengths by high altitude dayside clouds, increases the albedo and thereby lowers the temperature of the planet \citep{parmentier2016}. Secondly, radiatively active clouds increase the scattering and absorption in the thermal wavelengths, which results in a greenhouse effect below the clouds. Additionally, including heat exchange from $H_2$ dissociation and recombination lead to less day-night temperature contrast \citep{Tan&Komacek2019}. Further, we implement a tracer based equilibrium cloud model which is simpler than the more sophisticated microphysical cloud scheme with an appropriate cloud particle size distributions by \citet{Powell2018}. Implementing a cloud scheme which considers the locally unique cloud particle size distribution in cloud formation processes may change to outcome of our study. Furthermore, longer simulation runs up to 50000 days like in \citet{Sainsbury2023MNRAS.524.1316S} may evolve an altered temperature structures and altered overturning circulations which affects the cloud structure. Finally, using different set of hydrodynamic equations may change the jet structure and the overturning circulation our results significantly \citep{mayne2019,deitrick2020,Noti2023MNRAS}

\section{Conclusions}
\label{ch:conclusions_UHJ_Tint}
Convection is inhibited in hot Jupiter atmospheres above the radiative-convective boundary due to their strong irradiation. We demonstrated that ultra hot Jupiters with warm interiors have strong static stability, and low convective fluxes. Therefore, the advective component is more dominant in shaping the cloud structure in ultra hot Jupiters with warm interiors than the convective mixing. To the contrary, we see ultra hot Jupiters with colder interiors show less static stability which lead to more convective adjustment and higher convective fluxes than those with hotter interiors. Stronger cooling of the interior is expected in the ultra hot Jupiters with colder interiors due to opacity inhomogeneities and convection. Ultra hot Jupiters with warmer interiors have higher vertical turbulent heat fluxes. In combination with the vertical advection of potential temperature (altered circulation cells) and weaker cooling due to larger opacities in the deep atmosphere, the turbulent heat fluxes inhibit convection in the deep atmosphere in the warmer cases. As a consequence of the lower temperatures and increased convective mixing, a global cloud layer forms in the deepest inversion in the cooler cases. Therefore, the vertical extent of the cloud structures are increased as the internal temperature is decreased. Our results show that vertical convective mixing becomes consistently higher than advective mixing in cold interiors of highly irradiated gas giants. In addition, strong vertical convective mixing may occur in isolated regions around and inside the Rossby gyres on the nightside. Therefore, we suggest models to include a vertical convective mixing scheme which quantifies additional mixing in the Rossby gyres more adequately. Overall, convective mixing globally plays a minor role in keeping cloud particles aloft in ultra hot Jupiters.

Future investigations may extend this study by adding radiatively active clouds like in \citet{Roman2021}, \citet{Komacek2022PatchyClouds} and \citet{Malsky2024ApJ}. Moreover, this study can be continued by including multiple cloud species like in \citep{Roman2021} and \citet{Malsky2024ApJ}. Alternatively, a microphysical cloud scheme like CARMA \citep{Gao2018,Powell2018} could be coupled to the GCM.

\begin{acknowledgements}
P. Noti and E.K.H. Lee are supported by the SNSF Ambizione Fellowship grant (\#193448), a scheme of the  Swiss National Science Foundation. Calculations were performed on UBELIX (\url{http://www.id.unibe.ch/hpc}), the HPC cluster at the University of Bern. We thank the IT Service Office (Ubelix cluster), the Physikalisches Institut and the Center for Space and Habitability at the Universit\"at of Bern for their services. Data and plots were processed and produced using PYTHON version 3.9 \citep{van1995python} and the community open-source PYTHON packages \emph{Bokeh} \citep{bokeh}, \emph{Matplotlib} \citep{hunter2007}, \emph{cartopy} \citep{cartopy1,cartopy2}, \emph{jupyter} \citep{kluyver2016}, \emph{NumPy} \citep{harris2020}, \emph{pandas} \citep{reback2020pandas}, \emph{SciPy} \citep{jones2001}, \emph{seaborn} \citep{waskom2021}, \emph{windspharm} \citep{dawson2016} and \emph{xarray} \citet{hoyer2017}. Parts of this work have been carried out within the framework of the NCCR PlanetS supported by the Swiss National Science Foundation under grants 51NF40\_182901 and 51NF40\_205606.

The 1D radiative-transfer, gCMCRT and various other source codes are available on GitHub: \url{https://github.com/ELeeAstro}. 
The input of the Exo-FMS GCM and the post-processings with gCMCRT are available on Zenodo, \href{https://zenodo.org/records/13254749}{DOI: 10.5281/zenodo.13254749}. All other data and code are available from the authors on a collaborative basis.
All other data is available upon request to the lead author.
\end{acknowledgements}

%
%
\bibliographystyle{aa} 
\bibliography{literature.bib} 

\begin{appendix} 

\section{Vertical heat flow}
\label{ch:vertical_heat_flow}

For quantifying the advection of potential temperature, we follow a similar approach as in \citet{Sainsbury2023MNRAS.524.1316S}.
The mean vertical enthalpy advection $F_{\mathcal{H}r}$ \lbrack Wm\textsuperscript{-2}\rbrack \:can be quantified as in \citet{Sainsbury2023MNRAS.524.1316S} as
\begin{equation}
F_{\mathcal{H}r}= \rho c_p T w,
	\label{eq:mean_vertical_enthalpy_advection}
\end{equation}
where $\rho$ \lbrack kgm\textsuperscript{-3}\rbrack \:is the density of the gas, $c_p$ \lbrack Jkg\textsuperscript{-1}K \textsuperscript{-1} \rbrack \:stands for the specific heat capacity, and $w$ \lbrack ms\textsuperscript{-1}\rbrack \: denotes the vertical wind speed.
In order to address varying grid point sizes, we take the surface integral of mean vertical enthalpy advection so that Equation \ref{eq:mean_vertical_enthalpy_advection} becomes
\begin{equation}
\frac{ Q_{r}}{\partial t}= \int_{0}^{2\pi}\int_{0}^{2\pi} \rho c_p T w R^2 \cos{\phi} \,d\phi\,d\lambda,
	\label{eq:vertical_heat_flow}
\end{equation}
where $\phi$ and $\lambda$ are the latitudinal and longitudinal angles, and $R$ \lbrack m\rbrack \:stands for the distance from the planetary center to the specific pressure surface. $R$ has the relation $R= r_p + z$, where $r_p$ \lbrack m\rbrack \: is the radius of reference surface and $z$ \lbrack m\rbrack \: the height from the reference surface to the specific pressure surface. As most planets have $r_p \gg z$, we ignore $z$.

\section{Richardson number}
\label{ch:richardson_number}

Many studies rely on Miles’ instability condition if the Richardson number is lower than the critical value $Ri_g<Ri_c$ \citep[$Ri_c$ between 0.25 and 1]{Miles1961JFM....10..496M,Abarbanel1984PhRvL..52.2352A} to estimate the stability of the flow and the presence of turbulences \citep[see]{Egerer2023AMT....16.2297E,Ostrovsky2024NPGeo..31..219O}. The gradient Richardson number $Ri_g$ describes the ratio of the buoyancy $N$ \lbrack s\textsuperscript{-1}\rbrack \:(the Brunt–V\"ais\"al\"a frequency) to the wind shear $S$ \lbrack s\textsuperscript{-1}\rbrack \:as
\begin{equation}
Ri_g=\frac{N^2}{S^2}= \frac{g}{T_v}\frac{\dfrac{\partial \theta_v}{\partial z}}{\Bigg(\dfrac{\partial v_{h}}{\partial z}\Bigg)^2},
	\label{eq:richardson_number0}
\end{equation}
where $g$ \lbrack1ms\textsuperscript{-2}\rbrack \:is the gravity, $T_v$ \lbrack K\rbrack \:represents the virtual temperature, $\theta_v$ \lbrack K\rbrack \:stands for the virtual potential temperature, $z$ \lbrack m\rbrack \:denotes the height, and $v_{h}$ \lbrack ms\textsuperscript{-1} \rbrack\:is the horizontal velocity.
The $Ri_c$ has been altered in several ocean and atmospheric studies \citep[see]{Egerer2023AMT....16.2297E}. Recent observations revealed that turbulences even occur at $Ri_g$ up to 10 and more \citep[see]{Ostrovsky2024NPGeo..31..219O}. Nevertheless, low $Ri_g$ still indicates the presence of turbulent flow. In this study, we made use of 3 thresholds for $Ri_c=$ 0.25, 1 or 10 as different uncertainty thresholds.

\section{Turbulent heat flux}
\label{ch:turbulent_heat_flux}

For quantifying the turbulent heat transfer, we follow \citet{youdin2010}. They expressed the turbulent heat flux $F_{eddy}$ \lbrack Wm \textsuperscript{-2} \rbrack \:as
\begin{equation}
F_{eddy}= -K_{zz} \rho g \bigg( 1 - \frac{\nabla}{\nabla_{ad}}\bigg),
	\label{eq:turbulent_heat_transfer}
\end{equation}
where $K_{zz}$ \lbrack m \textsuperscript{2} s\textsuperscript{-1}\rbrack \:denotes the  (thermal) eddy diffusion coefficient, $\rho$ \lbrack kgm\textsuperscript{-3}\rbrack \:is the density of the gas, $g$ \lbrack1ms\textsuperscript{-2}\rbrack \:represents the gravity, $\nabla$ \lbrack K\rbrack \:is the vertical lapse rate (temperature gradient), and \: $\nabla_{ad}$ \lbrack K\rbrack \:denotes the adiabatic lapse rate.

\end{appendix}

\end{document}